\documentclass[3p,times]{elsarticle}

\usepackage{ecrc}

\volume{00}

\firstpage{1}

\journalname{Nuclear Physics A}

\runauth{T. Sakaguchi et al.}

\jid{procs}

\jnltitlelogo{Nuclear Physics A}

\usepackage{amssymb}

\usepackage[figuresright]{rotating}

\begin{document}

\begin{frontmatter}



\dochead{}

\title{PHENIX photons and dileptons}


\author[alabel1]{Takao Sakaguchi, for the PHENIX Collaboration}

\address[alabel1]{Brookhaven National Laboratory, Physics Department}

\begin{abstract}
Electro-magnetic probes such as dileptons and photons are strong probes
to investigate the thermodynamical state of the early stages of collisions
since they leave the system unscathed. The PHENIX experiment has measured
both photons and dileptons in p+p, d+Au and Au+Au collisions. An excess
of dilepton yield over the expected hadronic contribution is seen in
0.2-0.8\,GeV/$c^2$ in Au+Au collisions, which is prominent in lower $p_T$
and most central. Direct photons are measured through their internal
conversion to electron pairs. We saw a large enhancement in Au+Au collisions
over p+p yield scaled by the number of binary collisions. It turned out
from the latest results on d+Au collisions that this enhancement is not
explainable by a nuclear effect.
\end{abstract}


\end{frontmatter}


\section{Introduction}
\label{Intro}
Many intriguing phenomena have been observed at RHIC where a hot and
dense matter is expected to be created. The large suppression of the
yield of the single hadrons at high transverse momentum ($p_T$) suggested
that the matter is so dense to stem fast partons with large $Q^2$ emerged
from the very early stage in the medium~\cite{Adcox:2004mh}. The large
elliptic flow of particles and its scaling in terms of kinetic energy of
the particles suggests that the system is locally in equilibrium at
as early as 0.3\,fm/c. Although the hadronic probes already
exhibited many interesting phenomena, since they are suffered from strong
interactions in later stages to some extent, observation of more direct and
penetrating probes have been desired.

\begin{figure}[htbp]
\begin{center}
\includegraphics[width=120mm]{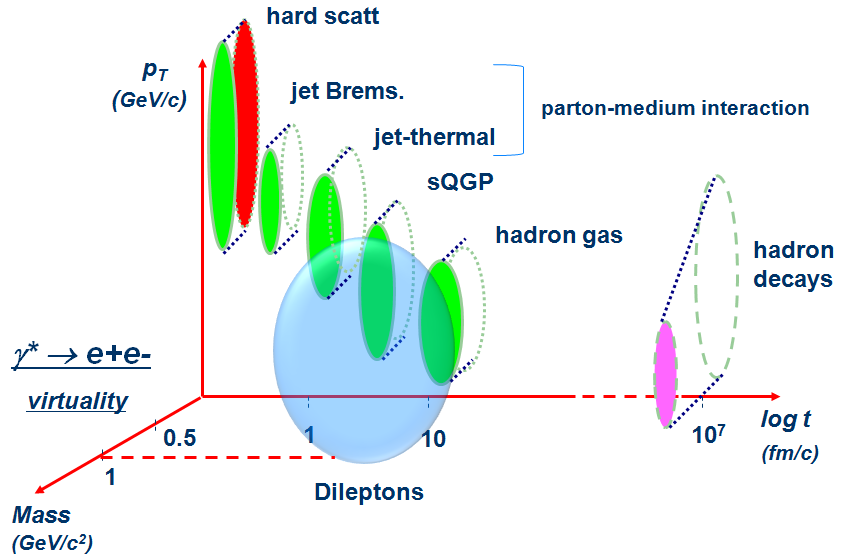}
\end{center}
\caption{Photons and dileptons in one coordinate with three degrees of freedom.}
\label{fig_photonmap}
\end{figure}

Electro-magnetic probes such as photons or dileptons are ideal in this sense,
since they interact with medium or other particles only electro-magnetically,
once produced~\cite{Stankus:2005eq}. Therefore, these probes are of interest
already from the beginning of the history of the heavy ion collisions.
At the leading order, the production processes of photons are
annihilation ($q\bar{q}\rightarrow\gamma g$) and Compton scattering
($qg \rightarrow \gamma q$). Their yields are proportional to
$\alpha\alpha_{s}$, which are $\sim$40 times lower than the ones from
strong interaction. The dilepton production is mainly from annihilation
of quark and anti-quark ($q\bar{q}\rightarrow\gamma^{*}$), and the
yield is proportional to $\alpha^{2}$. Except for some difference in
production processes, photons and dileptons can be
treated as same observables. A theoretical calculation tells that photons
and dileptons share similar emission sources~\cite{Turbide:2003si}.
Fig~\ref{fig_photonmap} shows a mapping of photons and dileptons in one
coordinate.

PHENIX has measured dileptons and photons since the RHIC has started its
running in 2000. In this paper, we show the most recent results on dileptons
and photons in p+p, d+Au and Au+Au collisions at $\sqrt{s_{NN}}$=200\,GeV.

\section{Dilepton analysis}
The PHENIX detector has a capability of measuring momentum of charged
particles with high accuracy and of identifying electrons with high
efficiency and strong rejection power of hadrons~\cite{Adcox:2003zm}.
In measuring dileptons, we have a huge combinatorial background $e^+e^-$
arising from random combination of electrons from photons converted at
beam pipe, Dalitz decays of $\pi^0$ or $\eta$. We constructed combinatorial
background by combining electrons or positrons from different events, and
subtracted them from foreground mass distributions.

The subtracted distributions still include unphysical correlated pairs
from back-to-back jets. We estimated the contribution using PYTHIA event
generator followed by a detector simulation, and subtracted it off from
the invariant mass distribution.  The signal to background ratio is
$\sim10^{-2}$ in total. The resulting distribution is corrected for
efficiencies of single electrons, and compared to various hadronic
contributions.

\section{Low mass and low $p_T$ dileptons in Au+Au collisions}
\begin{figure}[htbp]
\begin{minipage}{55mm}
\begin{center}
\includegraphics[width=51mm]{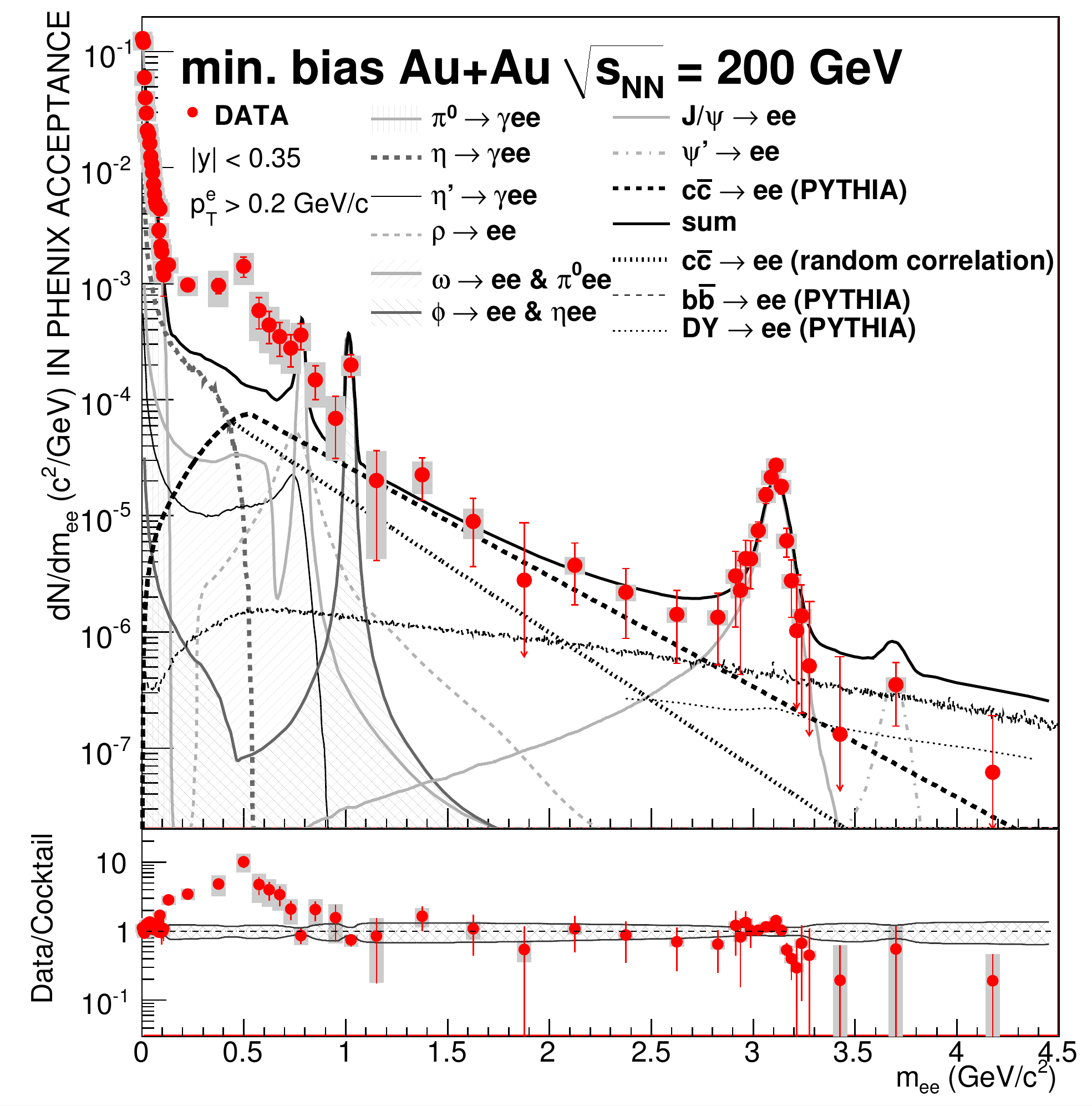}
\end{center}
\end{minipage}
\begin{minipage}{50mm}
\begin{center}
\includegraphics[width=48mm]{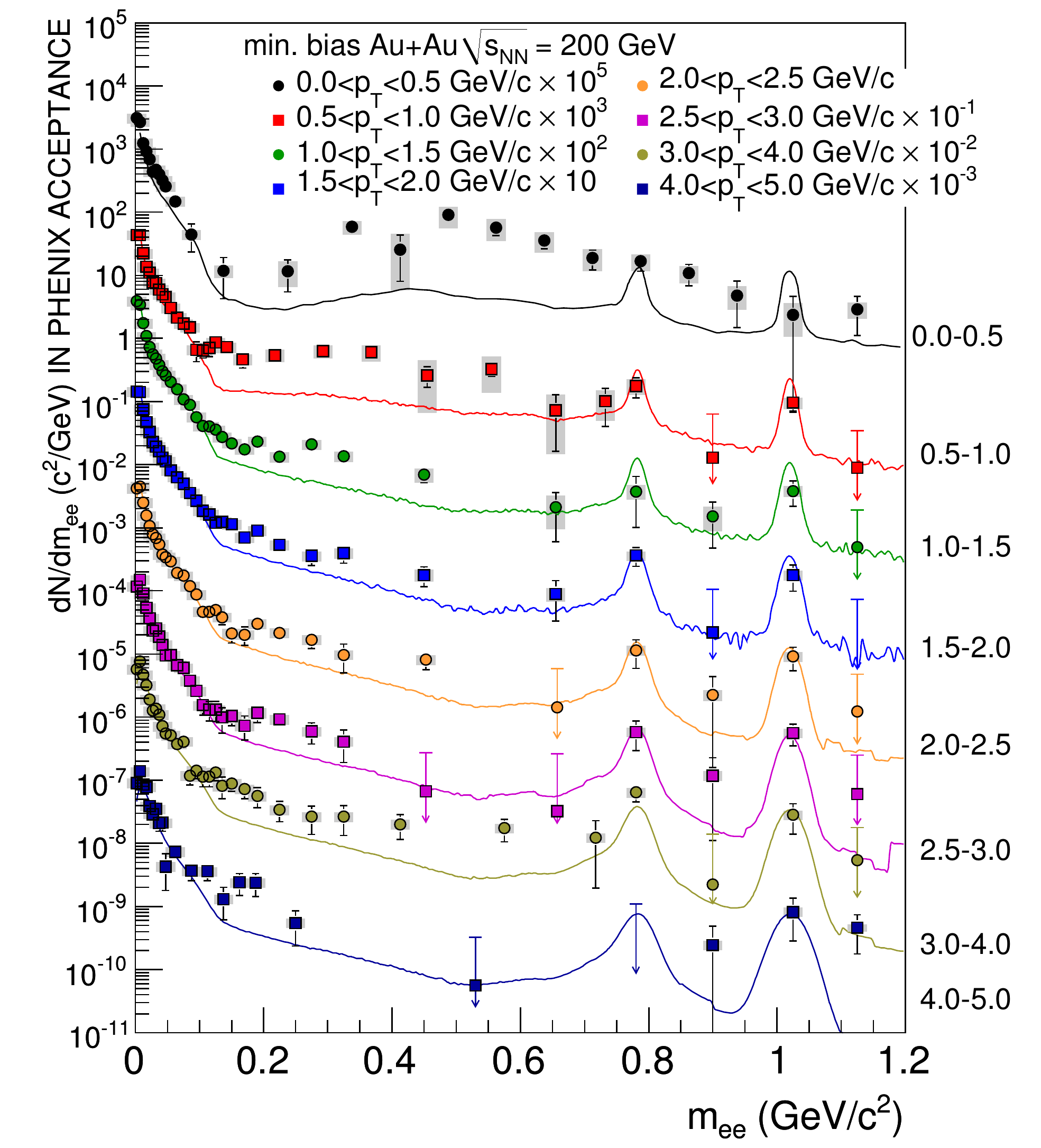}
\end{center}
\end{minipage}
\begin{minipage}{55mm}
\begin{center}
\includegraphics[width=51mm]{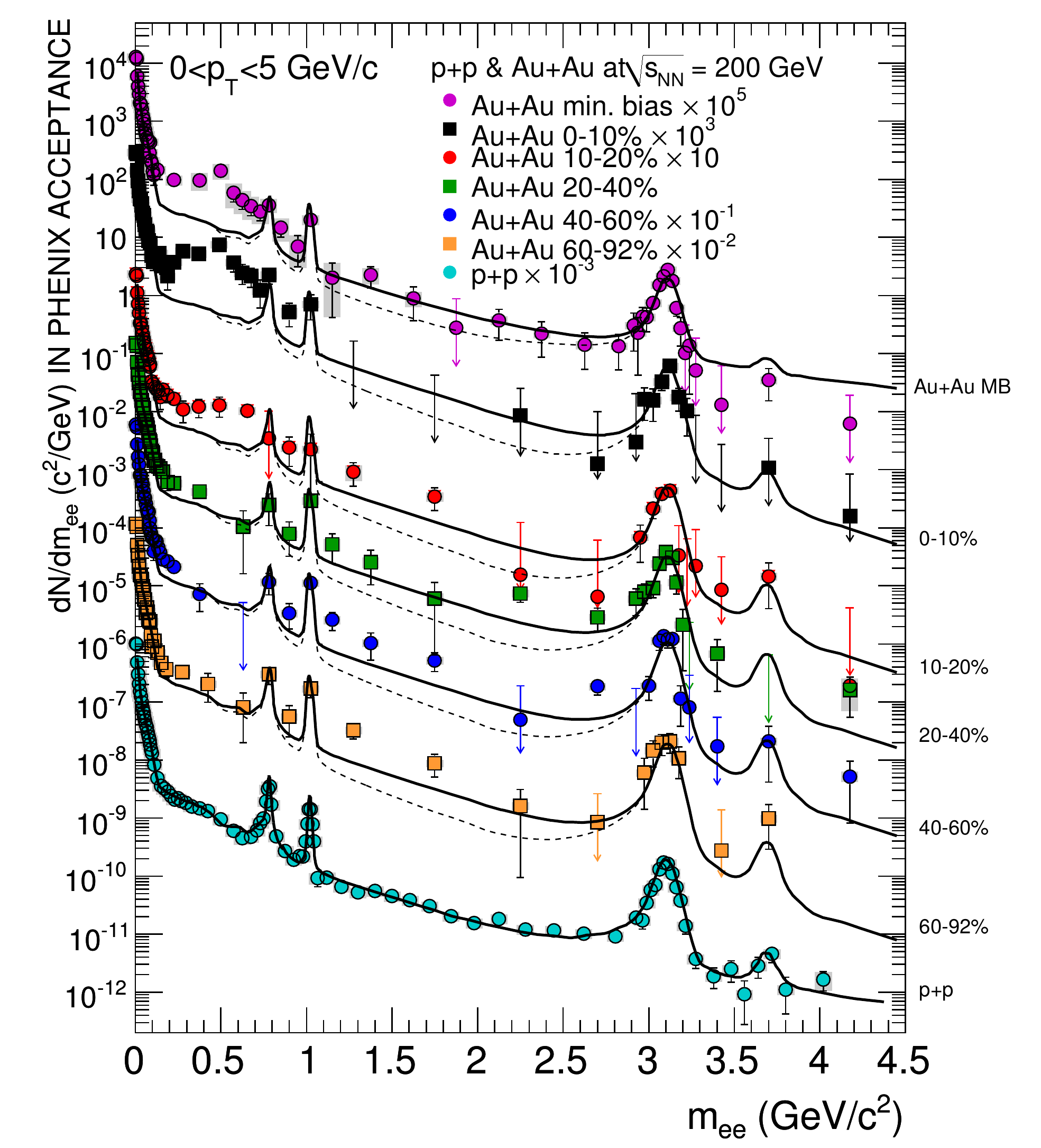}
\end{center}
\end{minipage}
\caption{(a, left) Invariant mass distributions in Minimum bias Au+Au events with hadronic cocktail calculation, and the ones in (b, middle) various $p_T$ regions and in (c, right) centrality classes.}
\label{fig_MinbDilep}
\end{figure}

\begin{figure}[htbp]
\begin{minipage}{75mm}
\begin{center}
\includegraphics[width=70mm]{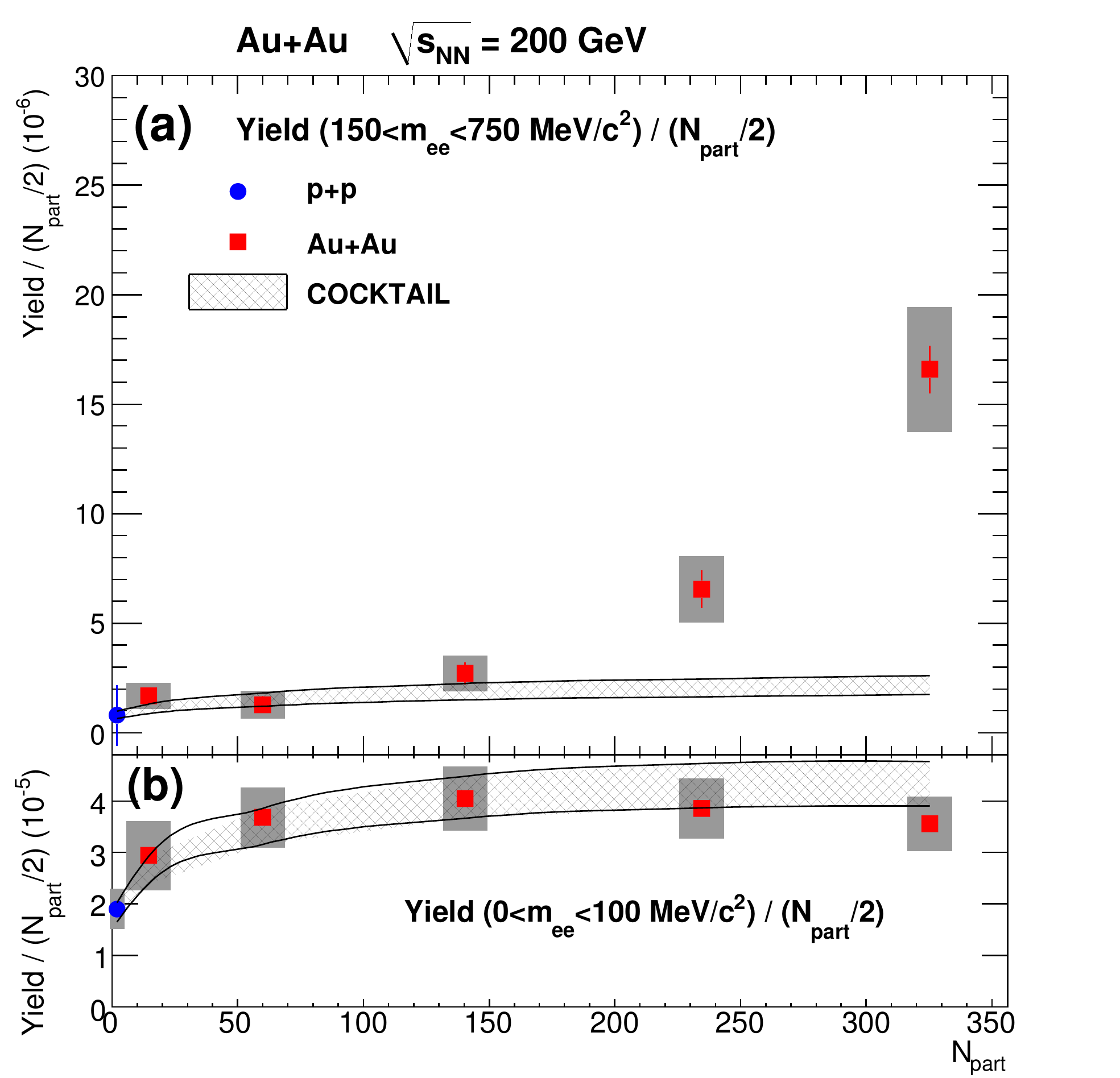}
\end{center}
\end{minipage}
\begin{minipage}{85mm}
\begin{center}
\includegraphics[width=80mm]{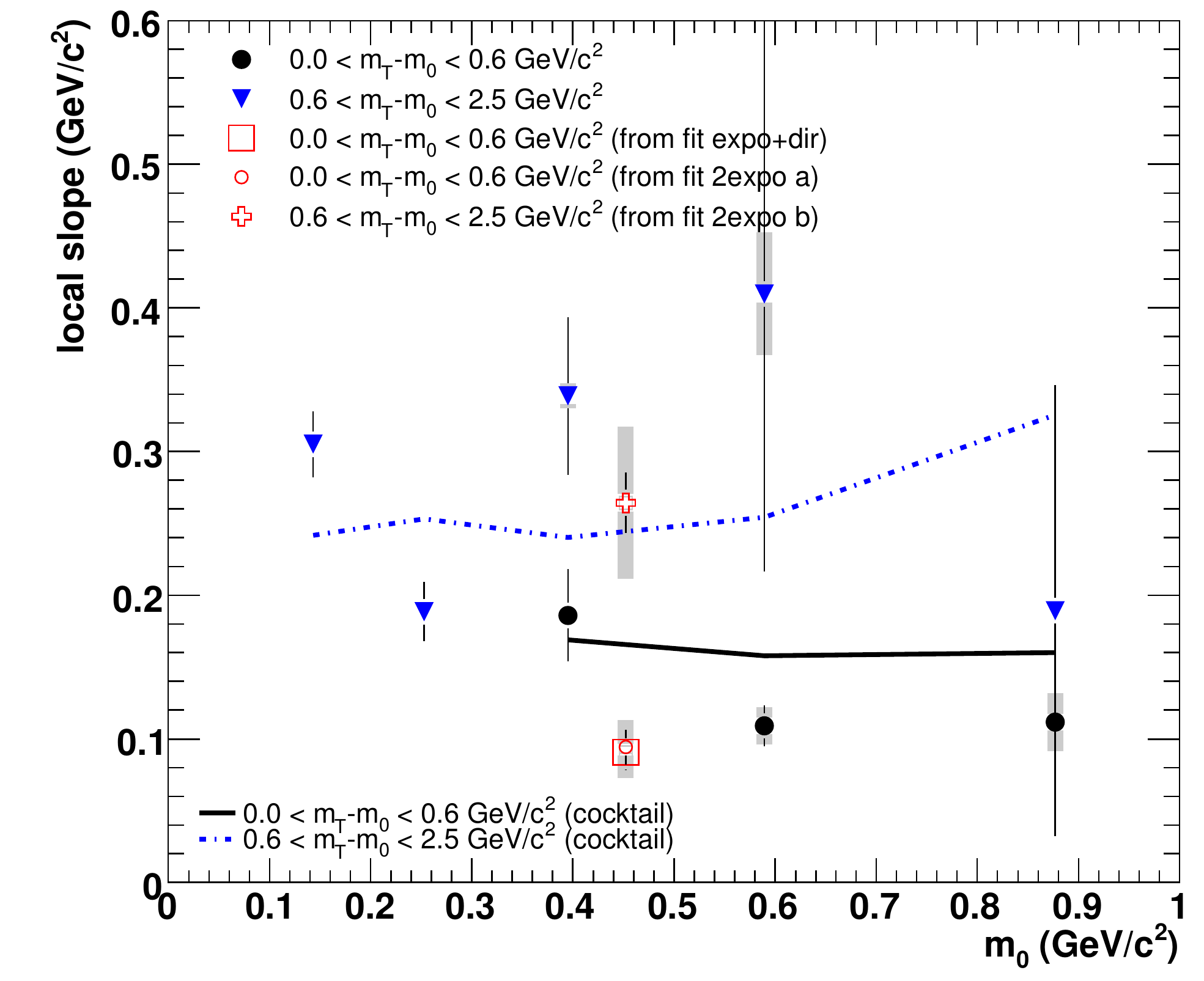}
\end{center}
\end{minipage}
\caption{(a, left) Yield and (b, right) slope parameters of dileptons in low mass and low $p_T$ region. For slope parameters, solid points are calculated from truncated means of the mass distributions, while open points are from exponential fits to the distributions.}
\label{fig_sumDilep}
\end{figure}

Fig~\ref{fig_MinbDilep}(a) shows the invariant mass distribution of electron
pairs measured in Au+Au collisions at
$\sqrt{s_{NN}}$=200GeV~\cite{Adare:2009qk}. The data is
compared with electron pairs from various hardonic sources calculated using
a Monte Carlo and filtered by the PHENIX acceptance. A large excess is seen
in 0.2-0.8\,GeV/$c^2$ in Minimum bias Au+Au collisions. In order to study
this excess in detail, we divided the data set into various $p_T$, mass and
centrality bins.
Fig.~\ref{fig_MinbDilep}(b) and (c) show the mass spectra in various $p_T$
and centrality bins, respectively.
The excess dies out at higher $p_T$ and lower centralities, which suggests that
the excess is contributed from a thermal source.
We made summary plots for the detail studies as shown in
Fig.~\ref{fig_sumDilep}(a) and (b).
As we already saw in the invariant mass distributions, the yield of excess
region increases as collisions become more central. The slope parameters are
low for low $m_{T}-m_{0}$ and high for high $m_{T}-m_{0}$, implying that
the local slopes are roughly proportional to the energy of photons emitted.
This is another evidence of that they are from thermal sources.

\section{Low mass and high $p_T$ dileptons in Au+Au and p+p}
\begin{figure}[htbp]
\begin{minipage}{80mm}
\begin{center}
\includegraphics[width=80mm]{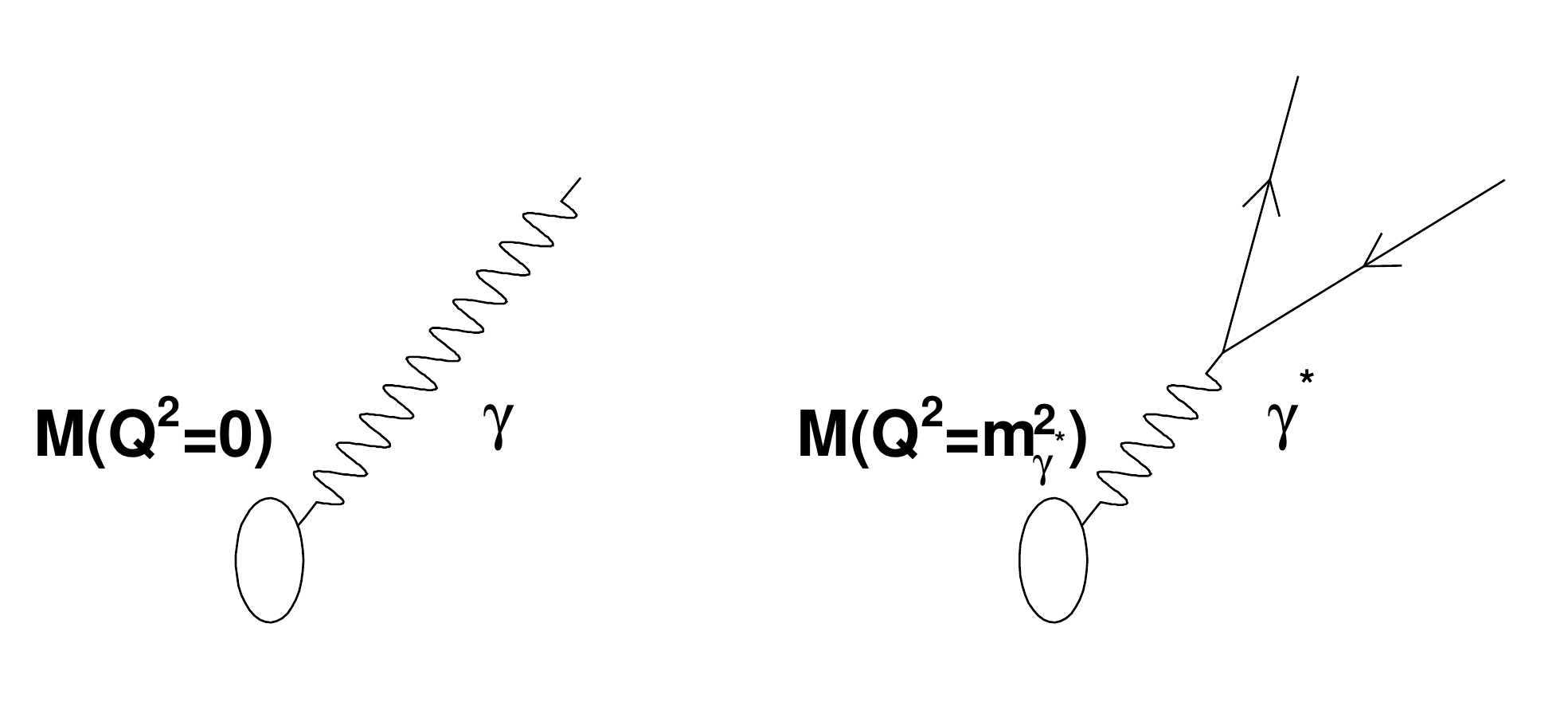}
\end{center}
\end{minipage}
\begin{minipage}{80mm}
\begin{center}
\includegraphics[width=80mm]{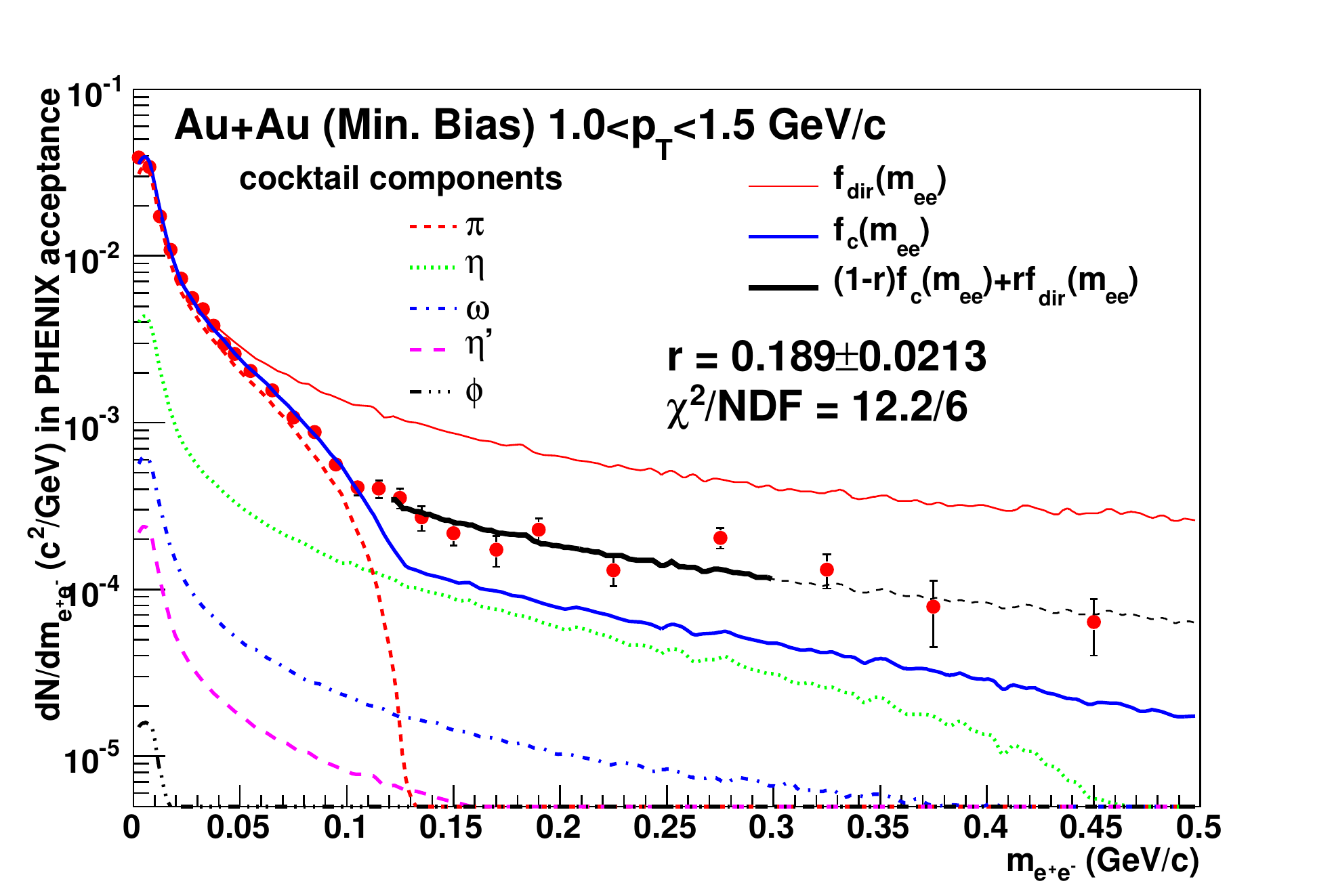}
\end{center}
\end{minipage}
\caption{(a, left ) Diagram of real photons and internal conversion of photons. (b, right) Fit to data with direct photon and hadronic cocktail distributions.}
\label{fig_incon}
\end{figure}

\begin{figure}[htbp]
\begin{minipage}{83mm}
\begin{center}
\includegraphics[width=78mm]{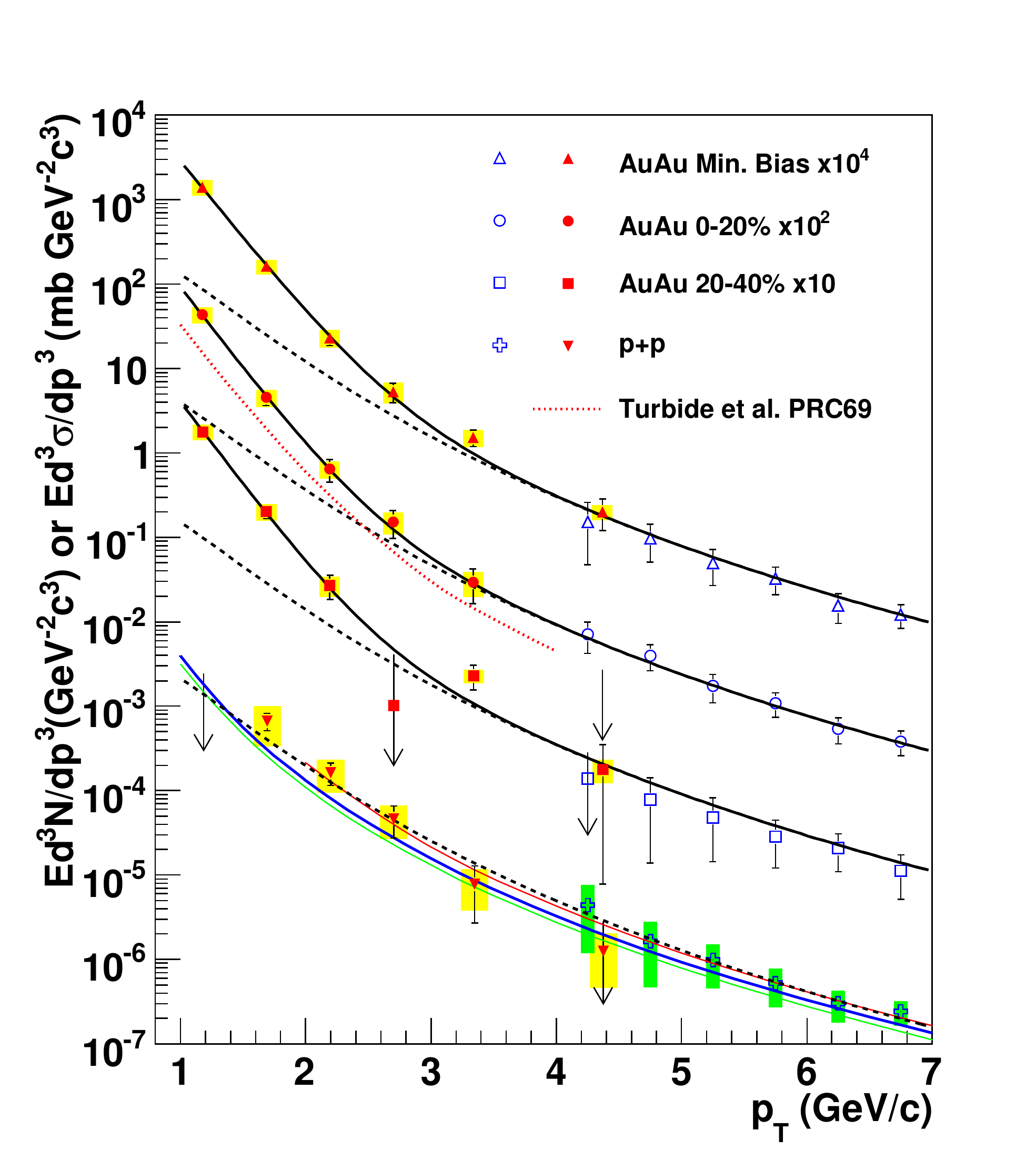}
\end{center}
\end{minipage}
\begin{minipage}{77mm}
\begin{center}
\includegraphics[width=72mm]{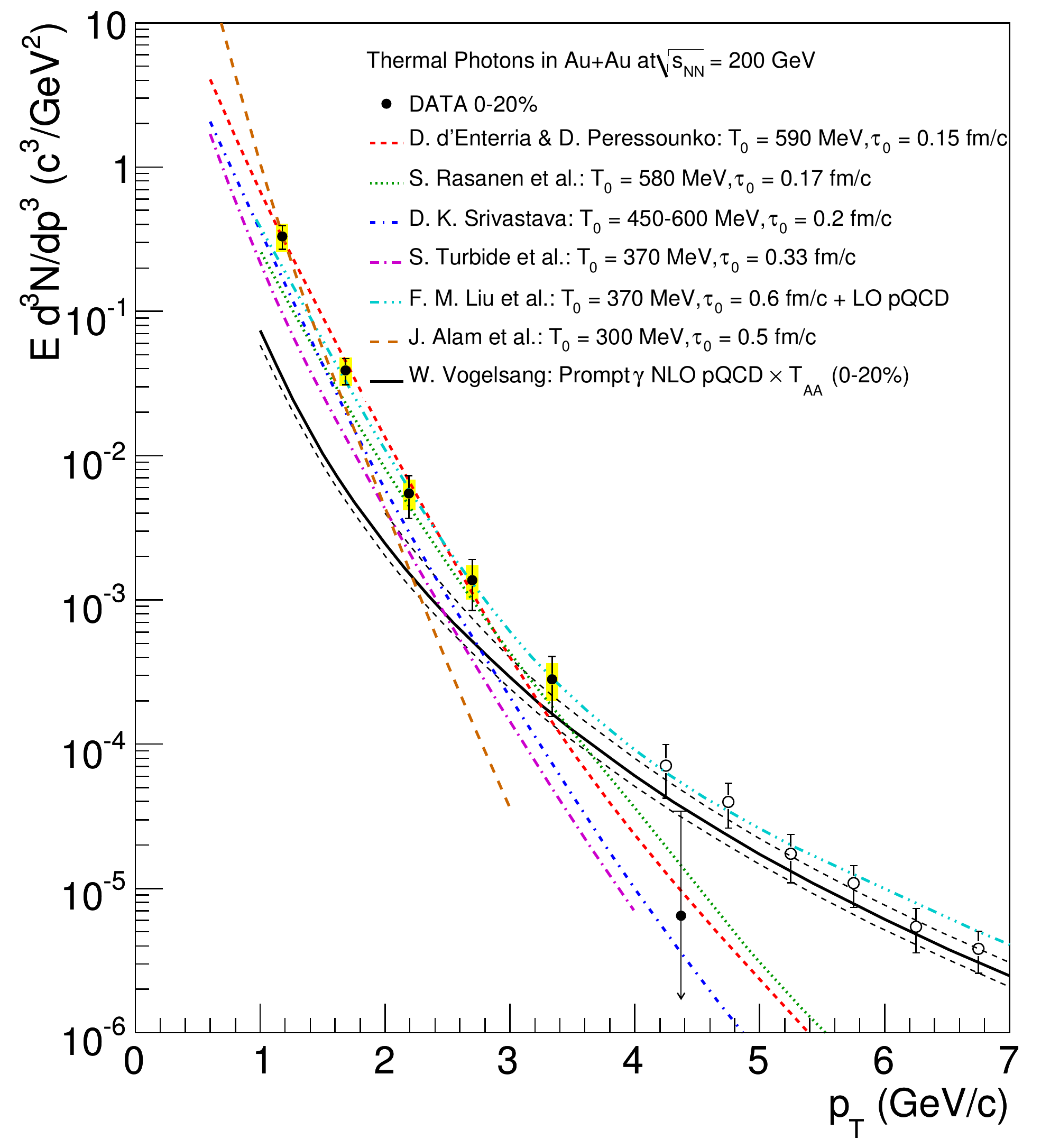}
\end{center}
\end{minipage}
\caption{(a, left) Direct photon yields in p+p and Au+Au collisions. (b, right) Comparison of the yield in 0-20\,\% Au+Au collisions with various theoretical models.}
\label{fig_dirphotonAuAu}
\end{figure}

Turning the region of interest to higher $p_T$, where $p_T>>M$,
the yield of dileptons are considered to be dominated by internal conversion
of real photons as depicted in Fig~\ref{fig_incon}(a).
Taking this advantage, we performed direct photon measurement through dilepton
measurement~\cite{Adare:2008fqa}. The invariant mass distribution of Dalitz
decay of $\pi^0$, $\eta$ and direct photons can be calculated using
Kroll-Wada formula as shown below.
\[ \frac{1}{N_{\gamma}} \frac{dN_{ee}}{dm_{ee}} = \frac{2\alpha}{3\pi}\sqrt{1-\frac{4m_{e}^{2}}{m_{ee}^{2}}}\left(1+\frac{2m_{e}^{2}}{m_{ee}^{2}}\right) \frac{1}{m_{ee}}|F(m_{ee}^{2})|^{2} \left(1-\frac{m_{ee}^{2}}{M^{2}}\right)^{3} \]
Then, the real invariant mass distribution is fitted with a function of:
\[ F = (1-r) f_c + r f_d\]
where $f_c$ is the cocktail calculation, $f_d$ is the mass distribution
for direct photons, and $r$ is the free parameter in the fit. Using Kroll-Wada
formula, $r$ is converted to direct photon to inclusive photon ratio as follows:
\[ r = \frac{\gamma^{*}_{dir} (m_{ee}>0.15)}{\gamma^{*}_{inc}(m_{ee}>0.15)} \propto \
 \frac{\gamma^{*}_{dir} (m_{ee}\approx 0)}{\gamma^{*}_{inc}(m_{ee}\approx 0)} \
= \frac{\gamma_{dir}}{\gamma_{inc}} \equiv r_{\gamma} \]
Once $r_{\gamma}$ is obtained, the invariant yield of direct photons are
calculated as $\gamma_{inc} \times r_{\gamma}$. The procedure demonstrated
in Fig~\ref{fig_incon}(b) is for 1.0$<p_T<$1.5\,GeV/c. The dotted lines show
the contributions from various hadrons, the solid blue is the sum of these
contributions, and the solid red line show the distribution by direct photons
converted internally. The $r$ value is determined by the fit to the data.
The error of the fit corresponds to the statistical error.
We applied the procedure for various $p_T$s and centralities in p+p and Au+Au
collisions, and obtained the $p_T$ spectra as shown in
Fig~\ref{fig_dirphotonAuAu}(a).
The result is obtained from 810 million Minimum bias Au+Au events collected
in RHIC Year-4 run, and 1.5 billion minimum bias events and 270 million
high $p_T$ single electron trigger events collected in RHIC Year-5 p+p run.
The distributions are for 0-20\,\%, 20-40\,\% centrality and Minimum Bias
collisions in case of Au+Au collisions. The distributions were then fitted
with the p+p fit plus exponential function. We obtain slopes and dN/dy for
three centralities as shown in Table~\ref{tab1}.
\begin{table}
\label{tab1}
\begin{center}
\caption{dN/dy and slopes of direct photon $p_T$ distributions obtained from
the fit to the data with the p+p fit plus exponential function.}
\begin{tabular}{c|c|c|c}
\hline\hline
Centrality & dN/dy ($p_T>$1\,GeV/c) & Slope (MeV) & $\chi^2$/DOF \\
\hline\hline
0-20\,\%  & 1.50$\pm$0.23$\pm$0.35 & 221$\pm$19$\pm$19 & 4.7/4 \\
20-40\,\% & 0.65$\pm$0.08$\pm$0.15 & 217$\pm$18$\pm$16 & 5.0/4 \\
MinBias   & 0.49$\pm$0.05$\pm$0.11 & 233$\pm$14$\pm$19 & 3.2/4 \\
\hline\hline
\end{tabular}
\end{center}
\end{table}
There are many models in the market that have a wide range of initial
temperatures and thermalization times.
Fig~\ref{fig_dirphotonAuAu}(b) shows the comparison of the data with
the models. Since the initial temperature
and thermalization time are highly correlated, the comparison of data
and models does not help constraining two parameters to definite numbers.

\section{Low mass and high $p_T$ dileptons in d+Au}
\begin{figure}[htbp]
\begin{minipage}{80mm}
\begin{center}
\includegraphics[width=75mm]{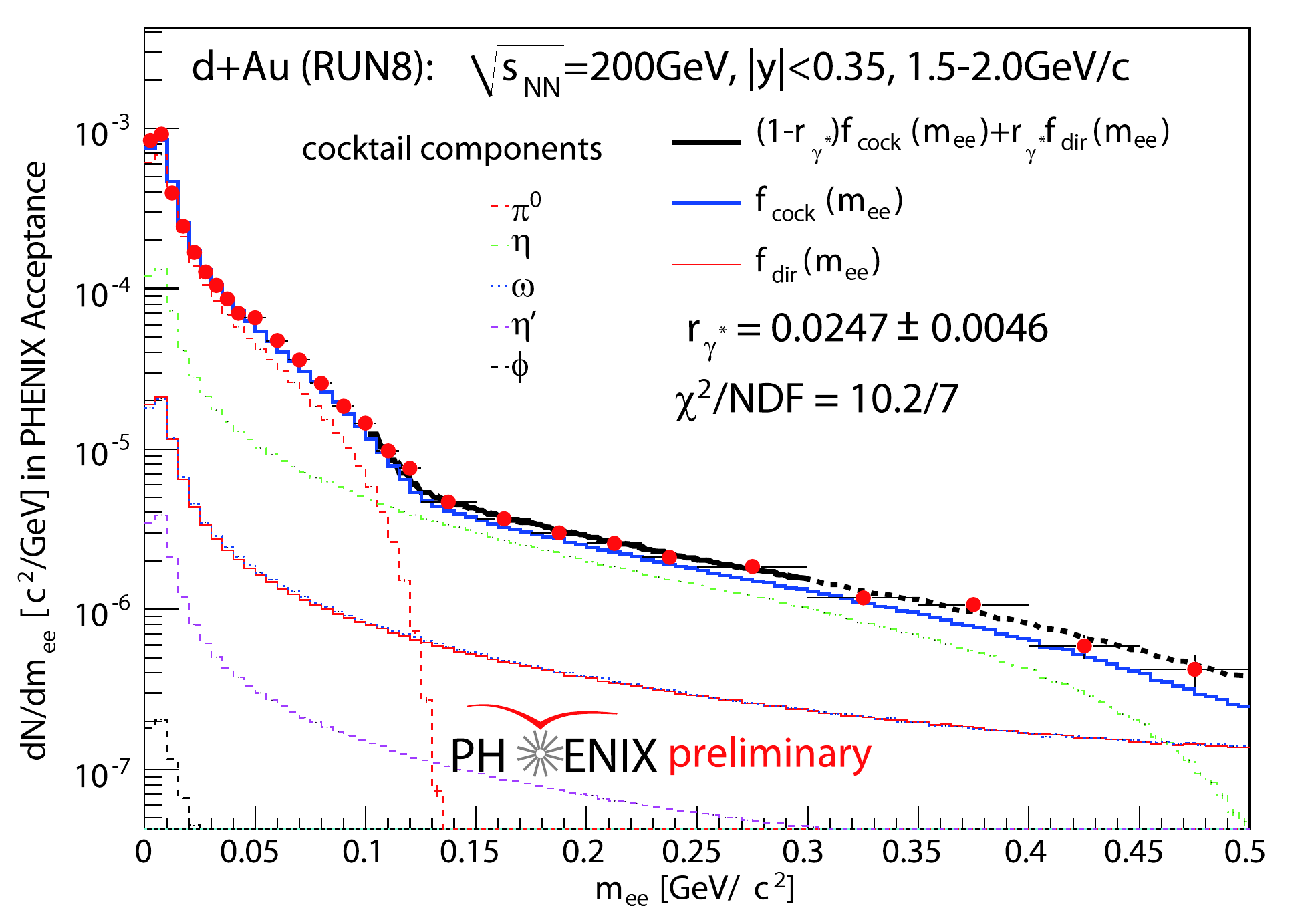}
\end{center}
\end{minipage}
\begin{minipage}{80mm}
\begin{center}
\includegraphics[width=80mm]{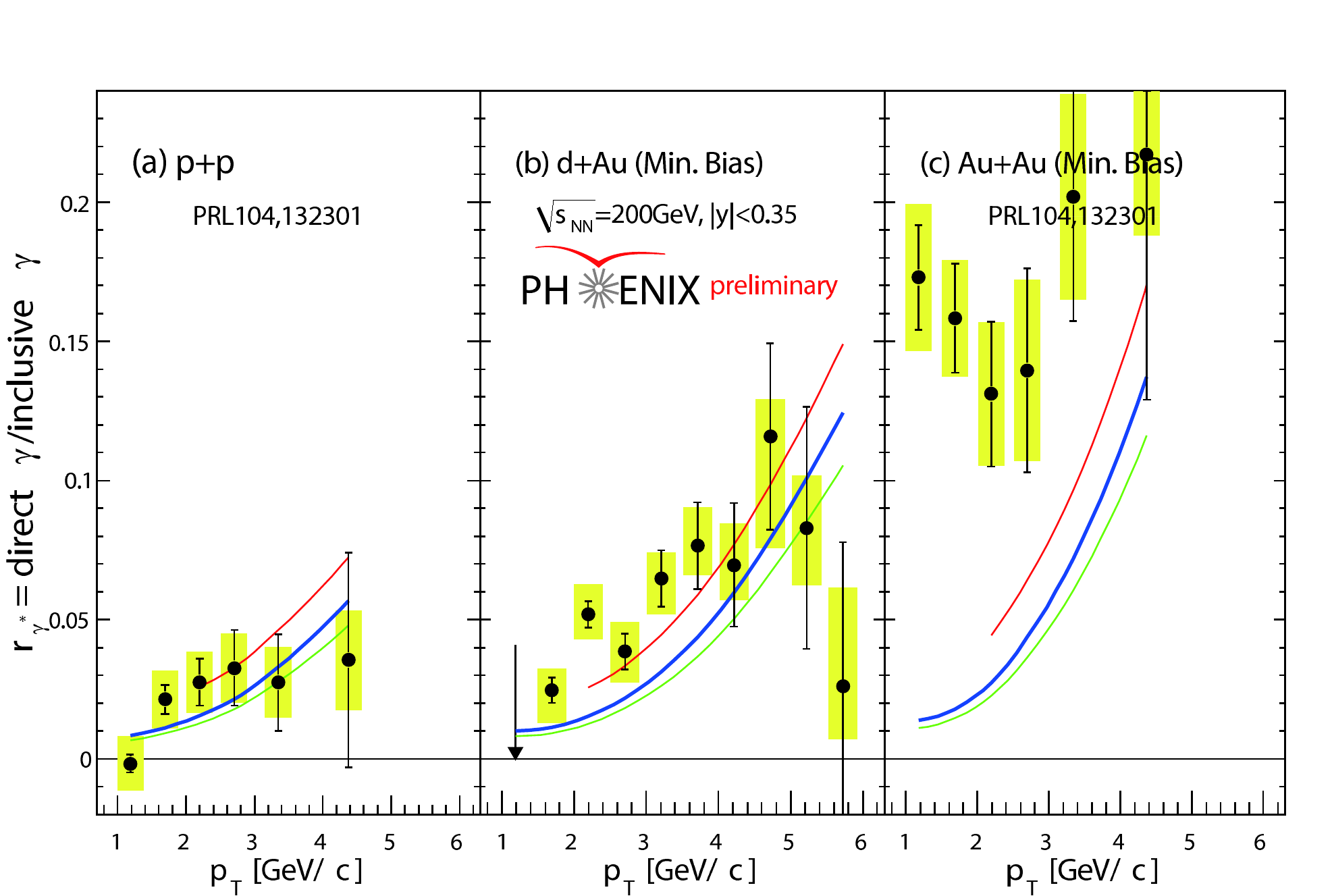}
\end{center}
\end{minipage}
\caption{(a, left) Invariant mass distribution together with the calculation for hadronic cocktail and direct photons. The procedure to obtain $r^{'}$ is described in analysis section. (b, right) Ratios of $\gamma_{dir}$ to $\gamma_{inc}$ for p+p, d+Au and Au+Au collisions.}
\label{fig_dAuphoton}
\end{figure}

\begin{figure}[htbp]
\begin{minipage}{80mm}
\begin{center}
\includegraphics[width=80mm]{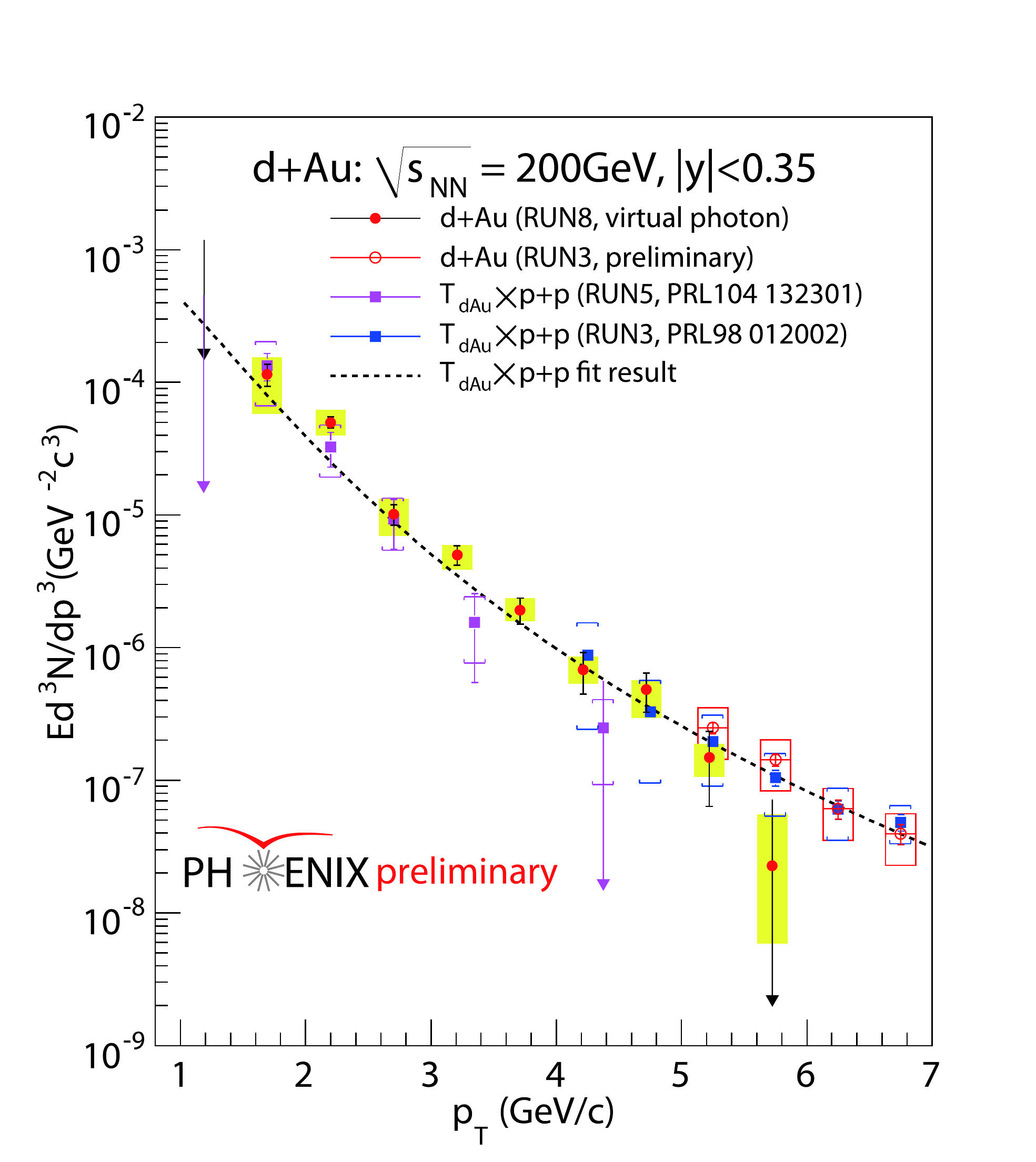}
\end{center}
\end{minipage}
\begin{minipage}{80mm}
\begin{center}
\includegraphics[width=80mm]{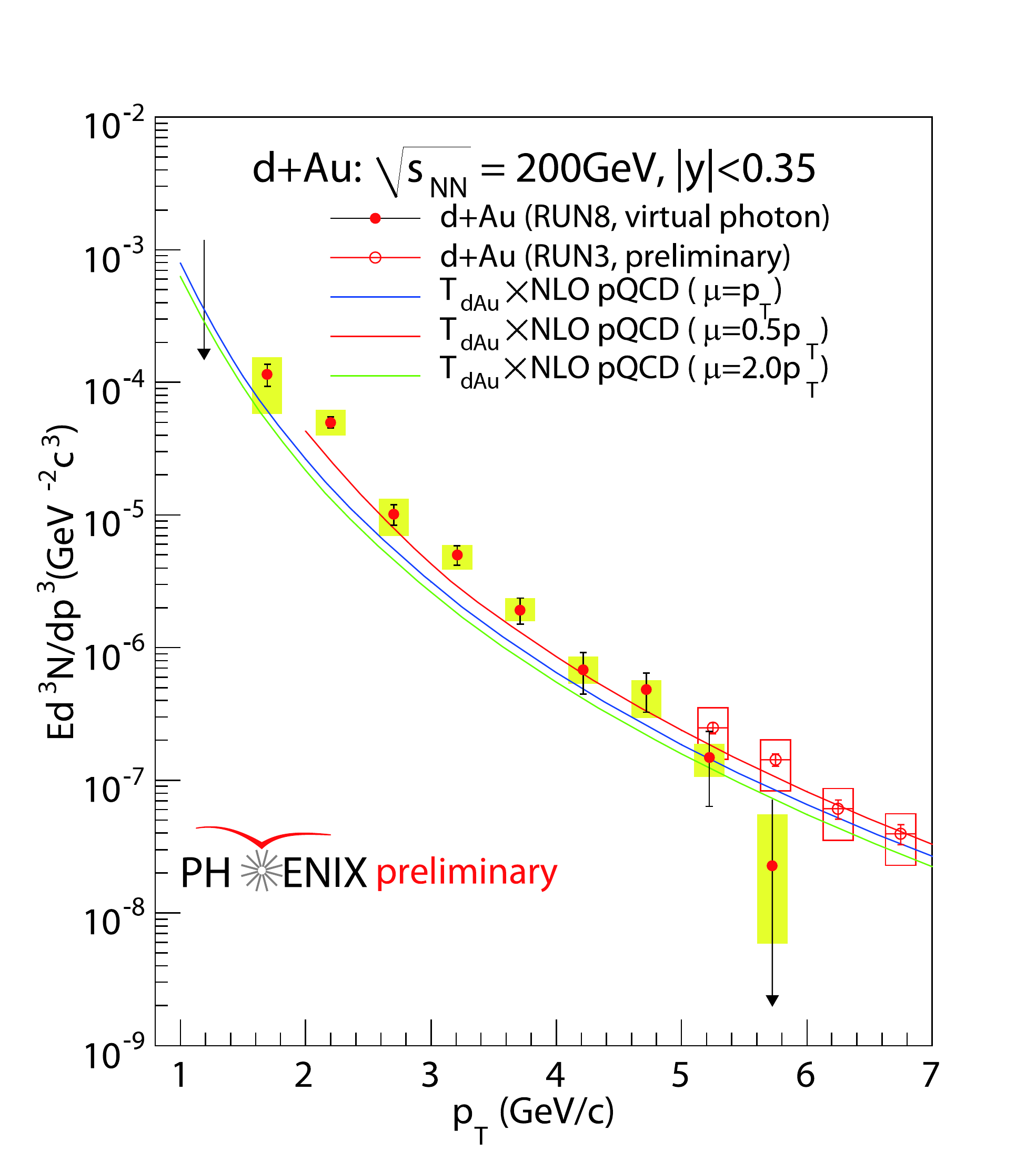}
\end{center}
\end{minipage}
\caption{(a, left) Direct photon yield in d+Au collisions from this analysis together with Run3 real photon result and p+p yield scaled by $N_{coll}$. (b, right) The same data with NLO pQCD calculation, instead of p+p yield.}
\label{fig_dAuSpec1}
\end{figure}

It has been a question whether or not the excess is purely from the source
that only exists in Au+Au collisions. A possible effect that increases the
the yield in nucleus collisions is $k_T$ broadening, or so-called Cronin
effect. In order to quantify the effect, we analyzed d+Au data from RHIC
Year-8 run with the same procedure.
Fig.~\ref{fig_dAuphoton}(a) shows the invariant mass distribution in d+Au
collisions for the lowest $p_T$ where the direct photon yield is significant.
After repeating over $p_T$ bins, we obtained the $r_{\gamma}$ as a function
of $p_T$. Fig~\ref{fig_dAuphoton}(b) shows the $r_{\gamma}$ in various
collision systems. The lines show the ratios calculated using the direct
photon contributions estimated by NLO pQCD prediction scaled by the number
of binary collisions ($N_{coll}$). Three lines show the predictions with
three mass scales ($\mu=p_T/2, p_T, 2p_T$).
The p+p ratio is very close to the NLO pQCD prediction, while Au+Au data
largely deviates in low $p_T$ from the prediction. The d+Au data shows a
moderate excess, but the excess is not as much as the one in Au+Au collisions.
From the ratios obtained, we calculated the direct photon spectra, again
following the procedure explained previously. Fig~\ref{fig_dAuSpec1}(a)
shows the direct photon spectra for d+Au Minimum bias events, together
with the real photon result from RHIC Year-3 run.
A comparison is made with direct photon spectra from p+p collisions scaled
by $N_{coll}$. Fig.~\ref{fig_dAuSpec1}(b) shows a comparison of data
with NLO pQCD calculation scaled by $N_{coll}$. In both cases, the yield in
d+Au collisions are higher than the one expected from p+p yield, suggesting
that the nuclear effect is seen in d+Au collisions.

\begin{figure}[htbp]
\begin{minipage}{80mm}
\begin{center}
\includegraphics[width=80mm]{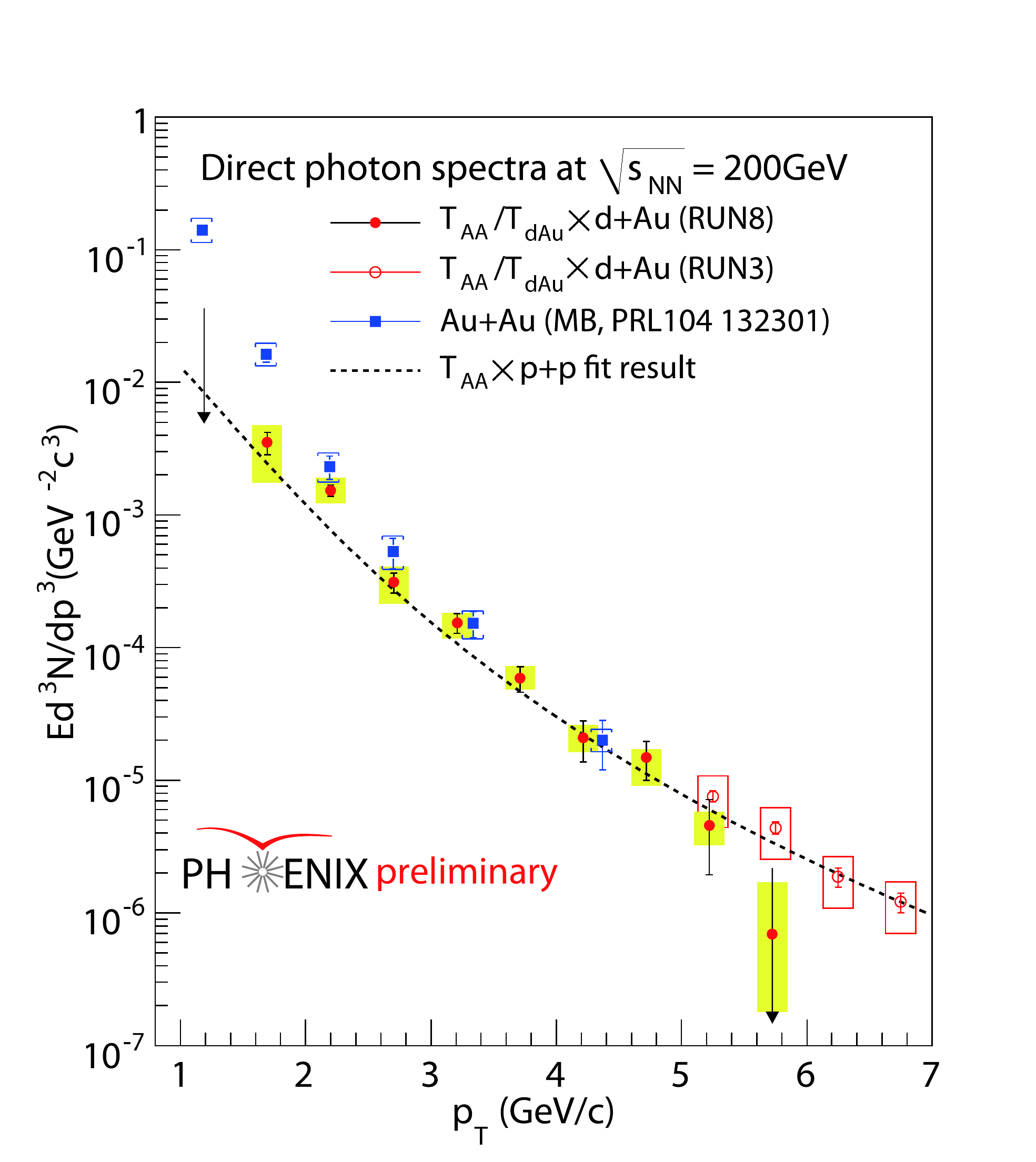}
\end{center}
\end{minipage}
\begin{minipage}{80mm}
\begin{center}
\includegraphics[width=70mm]{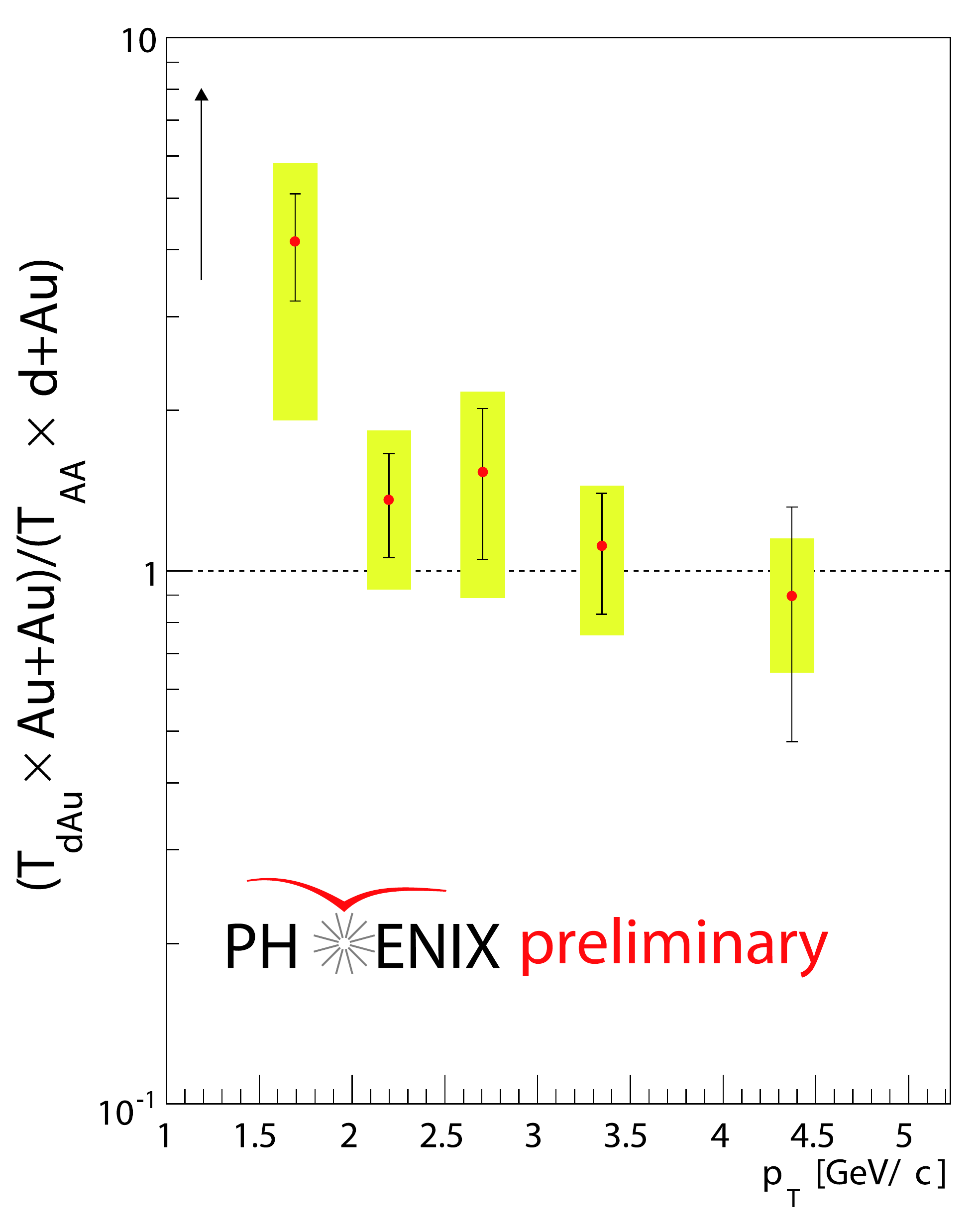}
\end{center}
\end{minipage}
\caption{(a, left) Direct photon yields in Au+Au and d+Au collisions scaled by the difference of $N_{coll}$ in both collision systems. (b, right) Ratio of Au+Au yield to d+Au yield scaled by $N_{coll}$.}
\label{fig_dAuSpec2}
\end{figure}

The ultimate interest in the series of the direct photon measurements
is whether or not the excess seen in Au+Au collisions can be explained by
cold nuclear effect (Cronin effect). In order to evaluate this, we compared
the Au+Au yield with the d+Au yield scaled by the difference of $N_{coll}$
as shown in Fig~\ref{fig_dAuSpec2}(a).
As seen in the plot, the Au+Au yield is higher than the one in d+Au
in $p_T<$2.5\,GeV/c, which is close to what is expected in
a literature~\cite{Turbide:2003si}.
When we divide Au+Au by $N_{coll}$-scaled d+Au data, we clearly see the
existence of an additional effect in Au+Au collisions
(Fig.~\ref{fig_dAuSpec2}(b)).

\begin{figure}[htbp]
\begin{minipage}{80mm}
\begin{center}
\includegraphics[width=80mm]{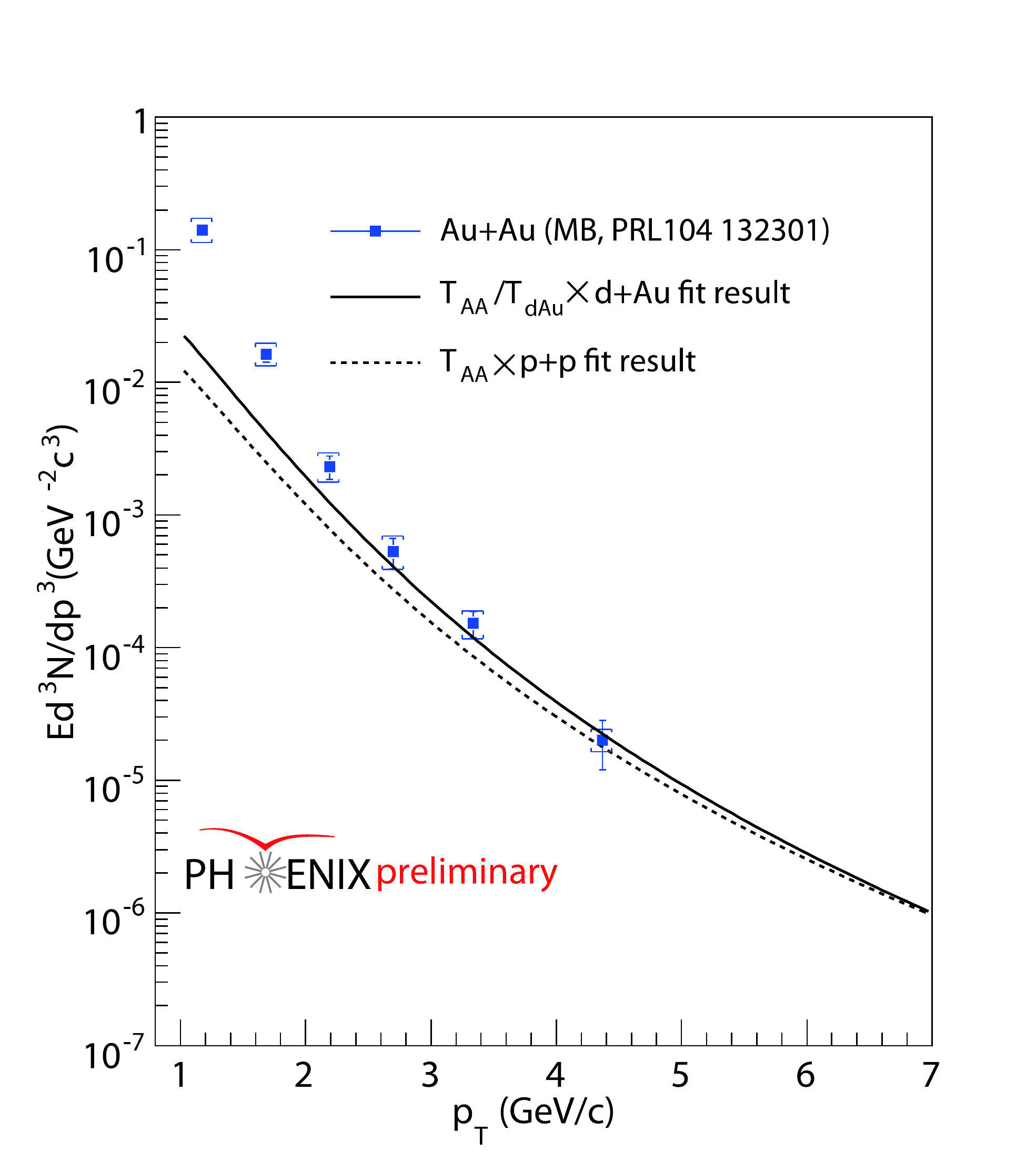}
\end{center}
\end{minipage}
\begin{minipage}{80mm}
\begin{center}
\includegraphics[width=80mm]{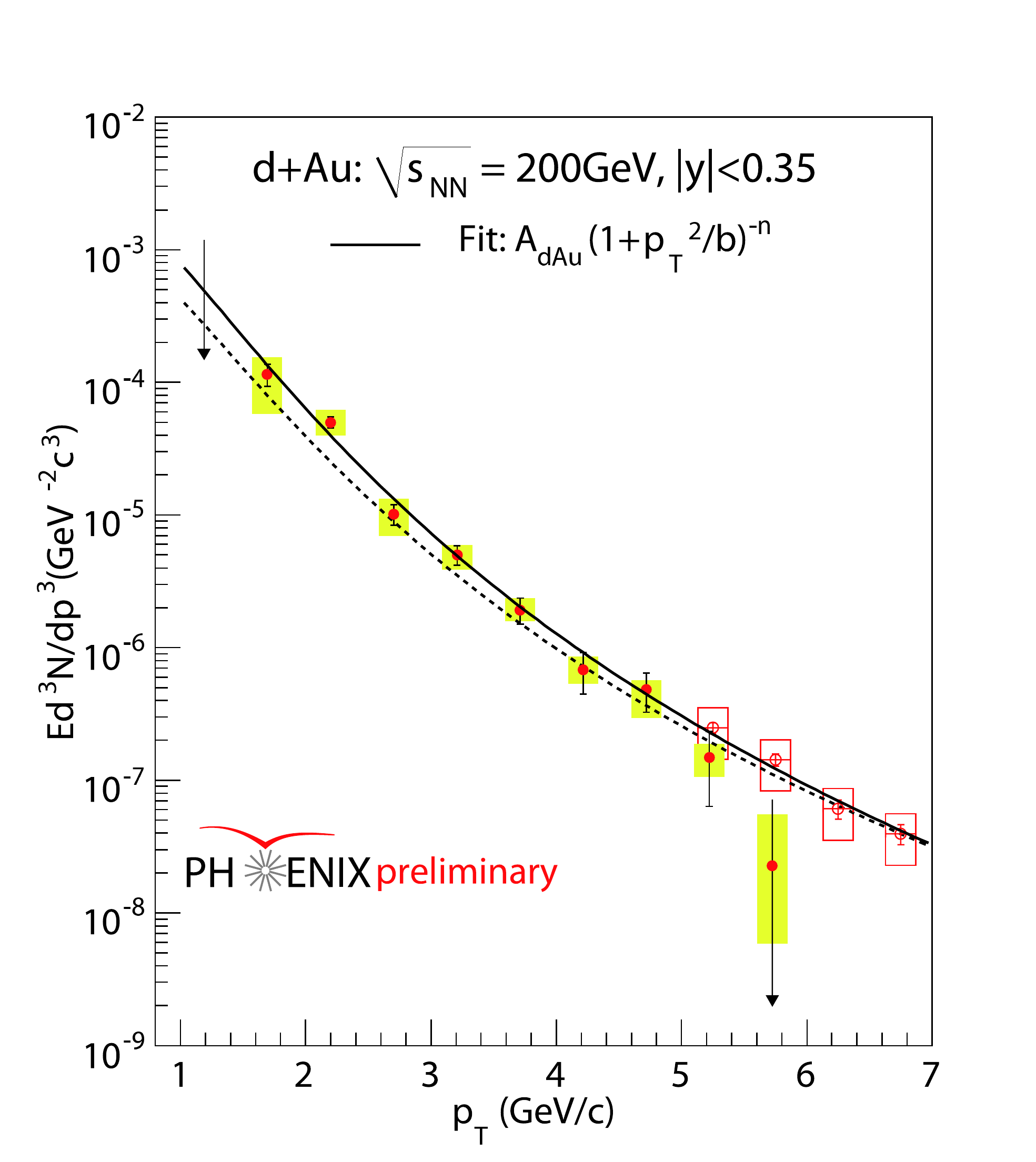}
\end{center}
\end{minipage}
\caption{(a, left) Fit to the d+Au invariant yield, and (b, right) comparison of the fit and Au+Au yield.}
\label{fig_dAuSpec3}
\end{figure}

Since both statistical and systematic uncertainties are still very
large in d+Au measurement, it is hard to exhibit a concrete quantitative
conclusion. However, in order to visually improve the result, we tried to
parameterize the d+Au points by using a fit function.
Fig~\ref{fig_dAuSpec3}(a) shows a fit to the d+Au points and
Fig~\ref{fig_dAuSpec3}(b) shows the fit scaled by $N_{coll}$ and Au+Au
points.
The excess in Au+Au for $p_T<$2.5GeV/c still looks to be definite compared
to the yield in d+Au, implying that the excess is caused by a source
created in hot dense medium.

\section{Toward better measurement of dileptons}
The statistical uncertainty and a part of systematic uncertainty
in dilepton measurement are governed by the a background arising
from random combinations of electrons from photons converted at
beam pipe and Dalitz decay of $\pi^0$ and $\eta$. If we could
tag and remove such electrons from foreground electron pair measurement,
the errors would be reduced significantly.

\begin{figure}[htbp]
\begin{minipage}{80mm}
\begin{center}
\includegraphics[width=80mm]{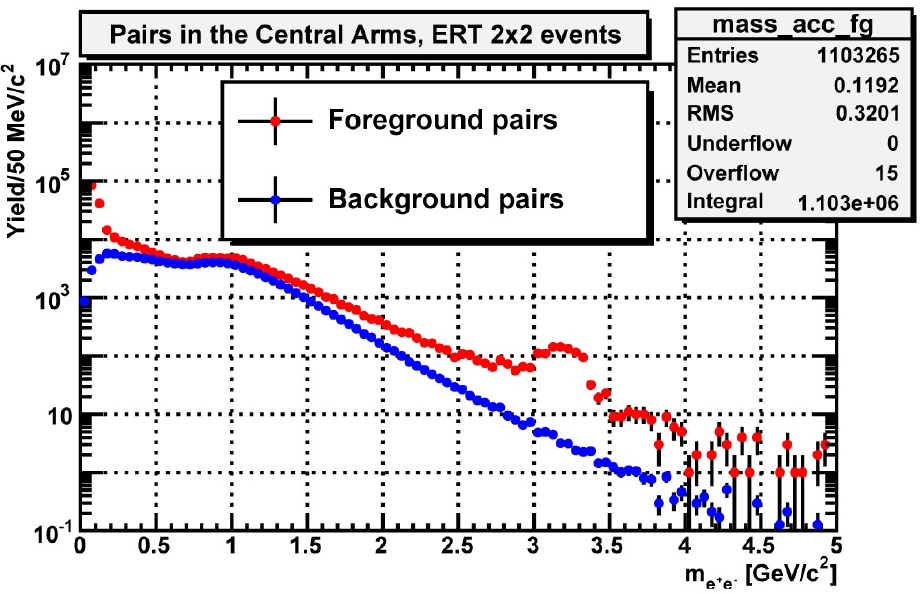}
\end{center}
\end{minipage}
\begin{minipage}{80mm}
\begin{center}
\includegraphics[width=80mm]{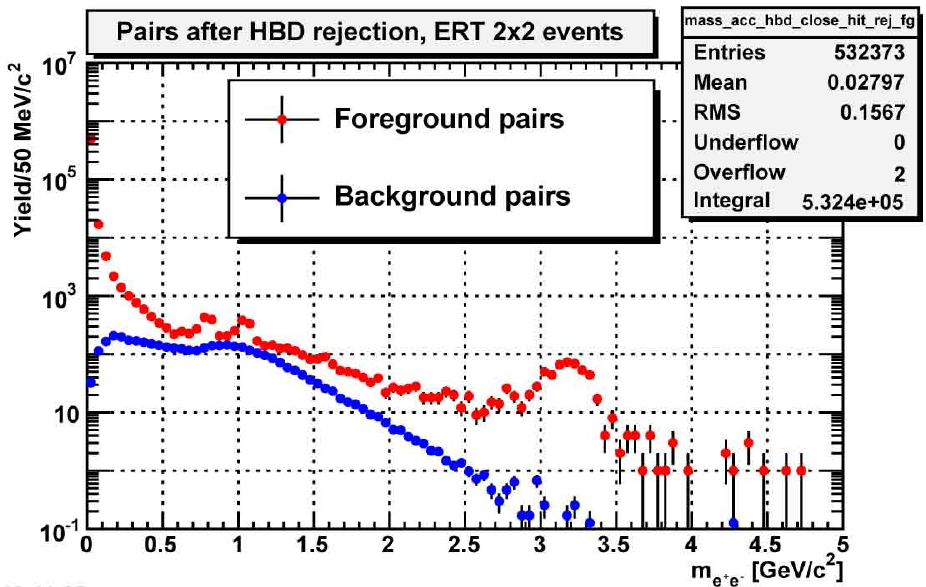}
\end{center}
\end{minipage}
\caption{Dilepton invariant mass spectra (a, left) without HBD matching and (b, right) with HBD matching and a charge cut.}
\label{fig_HBD}
\end{figure}

We developed a hadron blind detector (HBD) to realize this tagging. The
detector is windowless \v{C}erenkov type detector with CF4 gas in its
radiator. We operate the detector in magnetic field
free region in order to look for an opening angle of electron pairs.
The principle of operation is following; if we see a cluster that has
charge corresponding to the number of photo-electrons for single
electrons, we keep the track associated with the cluster. If we see a cluster
with double of the number of the photon-electrons, we reject the tract
associated with the cluster, assuming that this is contributed by two
electrons, which are likely by photon conversion or Dalitz decay. The detail of
the detector description and its performance can be found in~\cite{ref10}.
Fig.~\ref{fig_HBD} shows the invariant mass distributions with and without
using HBD information from Year-9 p+p run at $\sqrt{s}$=500GeV.
The signal to background ratios in $\phi$, $\omega$ and $\rho$ meson region
are significantly improved after applying a charge cut in HBD.
We are planning to also apply the charge cut in HBD in Au+Au collisions in
RHIC Year-10 run data. It is essential in Au+Au collisions because the particle
multiplicity is much higher than that in d+Au or p+p.

\section{Summary}
Electro-magnetic probes such as dileptons and photons are strong probes
to investigate the thermodynamical state of the early stages of collisions
since they leave the system unscathed. The PHENIX experiment has measured
both photons and dileptons in p+p, d+Au and Au+Au collisions. An excess
of dilepton yield over the expected hadronic contribution is seen in
0.2-0.8\,GeV/$c^2$ in Au+Au collisions, which is prominent in lower $p_T$
and most central. The invariant slopes are
low for low $m_{T}-m_{0}$ and high for high $m_{T}-m_{0}$, implying that
the local slopes are roughly proportional to the energy of photons emitted.
Direct photons are measured through their internal
conversion to electron pairs. We saw a large enhancement in Au+Au collisions
over p+p yield scaled by the number of binary collisions. It turned out
from the latest results on d+Au collisions that this enhancement is not
explainable by a nuclear effect.








\end{document}